\documentclass{optica-article}

\usepackage{opticajournal} 

\articletype{Research Article}

\usepackage{lineno}

\usepackage{geometry}
\usepackage{mhchem}
\newcommand\thefont{\expandafter\string\the\font}

\usepackage{acronym}
\acrodef{HHG}{high-harmonic generation}
\acrodef{TIPTOE}{tunneling ionization with a perturbation of the time-domain observation of an electric field}
\acrodef{TOF}{time-of-flight}
\acrodef{ATAS}{attosecond transient absorption spectroscopy}
\acrodef{CRIME}{complete reconstruction, using ionization yield modulation, of the electric field}
\acrodef{RDW}{resonant dispersive wave}
\acrodef{ADK}{Ammosov-Delone-Krainov}
\begin{document}

\title{Universal in-line waveform characterization using arbitrary non-linear responses}

\author{Chung Sum Leung, Joss Wiese, Katherine Brupbacher, and Hans Jakob Wörner\authormark{*}}
\address{Laboratorium für Physikalische Chemie, ETH Zurich, 8093 Zurich, Switzerland \\
\email{\authormark{*}hwoerner@ethz.ch}}

\begin{abstract*} 
Contemporary schemes for waveform-resolved characterization are constrained by setup-specific requirements, which severely limits their adaptability and fails to establish standard procedures for routine in-line diagnostic. This work reports a comprehensive experimental demonstration that relative yield measurements from a broad variety of media and nonlinear observables, combined with our family of open-source reconstruction algorithms (CRIME and lazyCRIME), allow for robust waveform retrieval with attosecond accuracy on a standard workstation in just minutes. We have further adapted this framework to multiple configurations---including non-invasive, simultaneous waveform characterization during an \ac{ATAS} experiment---showcasing the low-cost and non-intrusive nature of the new pulse characterization approach. Together, this work establishes an easy-to-implement universal characterization scheme for in-line diagnostic of ultrashort pulses that is readily accessible to the broader ultrafast science community.
\end{abstract*}


\section{Introduction}
Ultrafast spectroscopy continuously advances toward resolving ever-shorter timescales in physical systems. With a growing understanding of light–matter interactions on the attosecond time scale and the concomitant observation of subfemtosecond dynamics~\cite{Baykusheva:20,Uiberacker:07,Kluender:11,Kraus:15,Matselyukh:25}, waveform-resolved characterization of light sources creating ultrashort pulses is essential to resolve ultrafast molecular dynamics, to enable precise temporal control of the laser-electric field and, thereby, of the light-driven dynamics~\cite{Niederhausen:08,Hentschel:01,gaumnitz17a,Vismarra:24}.

A wide variety of established pulse characterization techniques exist, each with specific limitations. Spectral interference schemes such as SPIDER (spectral phase interferometry for direct electric field reconstruction) \cite{Wong:94} and MEFISTO (measurement of electric field by interferometric spectral phase observation) \cite{Amat-Roldan:05} rely on spectral continuity and cannot reconstruct pulses with disjoint spectral islands \cite{Keusters:03}; handling octave-spanning spectra often requires additional, setup-specific complexity \cite{Fan:16}. Related time-frequency methods (for example, FROG (frequency-resolved optical gating) \cite{Kane:93} and dispersion scan \cite{Miranda:12}) are constrained by the bandwidth of optics and detectors and, depending on geometry, may sacrifice temporal resolution to suppress unwanted interference between the gated response and the fundamental field \cite{Keller:22,Gallmann:00}. Attosecond streaking \cite{Goulielmakis:04,Itatani:02} provides direct access to the waveform-resolved structure but demands XUV attosecond sources and high photoelectron energy resolution, limiting routine applicability.

Recent developments in \acsu{TIPTOE} \cite{Park:18} provide a more accessible complementary route to previous established characterization techniques: waveform measurement and reconstruction from the delay-dependent modulation of a nonlinear response, driven by a strong field and perturbed by a weaker field. Due to their comparatively simple implementation, methods such as \ac{TIPTOE} and its close relatives have progressed rapidly, with demonstrations ranging from hundreds of femtoseconds down to ultrashort pulses such as solitons and \acp{RDW} \cite{Yeom:24,Heinzerling:25}, which have led to the generation of single-cycle pulses and even sub-cycle transients \cite{kopp25a,lanfaloni25a}. The technique has been adapted to a wide range of observables: ionization yield, fluorescence yield, high-harmonic yield, electric charge signals, photoconductive currents, across gas and condensed phase targets \cite{Saito:18,Awad:24,Schotz:22,Schiffrin:13}. Notably, Ref. \cite{Han:25} reported \ac{TIPTOE} based on acoustic signals recorded with a household microphone, and single-shot operation has been demonstrated in non-collinear geometries \cite{Yeom:24}, underscoring the versatility of the method. \ac{TIPTOE} has recently been generalized to measure the time-dependent polarization of ultrashort laser pulses \cite{Matselyukh:25b}.

Beyond the measurement scheme itself, \ac{TIPTOE}-based reconstruction has also advanced. Our previous work \cite{Wiese:24} introduced \acsu{CRIME}, which retrieves the waveform-resolved electric fields of both pulses involved on absolute scale, without intrinsic bounds on relative irradiance, pulse duration, or spectral range.

Here, we further advanced the paradigm of universal waveform characterization with a comprehensive experimental demonstration.  We show that waveform retrieval based on the TIPTOE scheme is widely independent of both the nonlinear response and the medium: integrated observables from disparate gases and detection channels yield equivalent waveform reconstruction. By integrating the total signal response without manually choosing regions of interest, we reduced the need for post-processing. Furthermore, leveraging our \ac{CRIME} algorithm and a reduced-input variant that we dub lazyCRIME, we demonstrate that waveform retrieval can generally proceed using model-of-choice up to the limit required for temporal reconstruction, allowing measurement and retrieval routines on a standard workstation in minutes. As a result, both \ac{TIPTOE}-based measurement and \ac{CRIME}-based reconstruction are decoupled from the specific medium and detector and are broadly applied under mild restrictions. Practically, this enables in-situ, non-intrusive, waveform-resolved characterization with high accuracy and robustness within minutes, establishing a general measurement archetype that can be deployed across diverse time-resolved spectroscopy setups. The source codes for \ac{CRIME} and lazyCRIME are openly available (see Data Availability Statement in \ref{sec:dataavail}).

\section{Measurement framework}
\begin{figure}
\centering\includegraphics[]{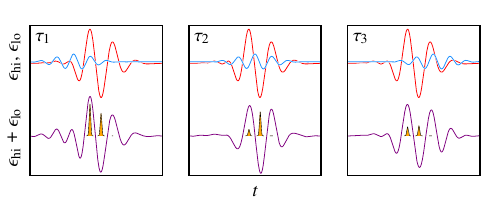}
\caption{\label{fig:scheme} Schematic representation of the \acs{TIPTOE} measurement scheme (adapted from \cite{Wiese:24}). At three different delays $\tau_1$, $\tau_2$ and $\tau_3$ between $\epsilon_{\text{hi}}$ (red) and $\epsilon_{\text{lo}}$ (blue), the strong field drives a general nonlinear response that is modulated by the weak field. The combined field (purple) produces a delay-dependent response (orange area).}
\end{figure}

The proposed pulse characterization framework is based on the established experimental procedure of \ac{TIPTOE}, in which the strong-field-ionization rate induced by a strong pulse, $\epsilon_\text{hi}$, is perturbed in a time-resolved manner by a weaker field, $\epsilon_\text{lo}$. Fig. \ref{fig:scheme} illustrates the concept for three different delays $\tau$ between $\epsilon_{\mathrm{hi}}$ and $\epsilon_{\mathrm{lo}}$. The time-integrated yield of ionization---or of any other nonlinear response---caused by the irradiation with $\epsilon_{\mathrm{hi}}$ and $\epsilon_{\mathrm{lo}}$ is related to the yield caused by $\epsilon_{\mathrm{hi}}$ alone. The resulting relative yield can then be used for waveform reconstruction with either \ac{CRIME} or lazyCRIME, depending on the desired information and experimental constraints. Together, these algorithms cover a wide range of experimental conditions with minimal compromise in reconstruction fidelity.

In this work, we performed back-to-back measurements and the corresponding waveform reconstructions from relative yields across gas-phase atomic and molecular targets and surveyed diverse nonlinear observables. Regardless of the observable---ionization yield, high-harmonic yield, fluorescence yield, and sound intensity---all relative yields exhibit the same temporal structure and thus enable high-fidelity waveform reconstruction. Consequently, we show that any nonlinear light-matter interaction can serve as measurement signal, provided its time-integrated magnitude can be recorded with adequate signal-to-noise ratio. Pairing with CRIME's reconstruction capability, our new medium-independent framework enables highly adaptable, routine waveform-resolved characterization with unprecedented accuracy, strong robustness, and rapid turnaround, thus fundamentally improving the accessibility of ultrashort pulse characterization in daily operation.

\section{Experimental implementation}\label{subsect:expschem}

All measurements were performed at a beamline that is commonly used for \ac{ATAS} experiments. To support an easy reproduction, all media and diagnostics were chosen such that they are commonly available in ultrafast spectroscopy laboratories. First, measurements of fluorescence yield and plasma sound intensity were conducted in ambient air, before the laser beam was coupled into the vacuum compartments of the beamline. This setup is described in \ref{subsub:ambient}. Second, measurements using ion, high-harmonic and fluorescence detection were performed in situ, in the vacuum compartment of the beamline, which is delineated in \ref{subsub:vacuum}. We would like to especially highlight the in-situ fluorescence detection, because it offers a completely nonintrusive scheme for waveform retrieval that could, in principle, be run \emph{simultaneously} with \ac{ATAS} measurements.

\begin{figure}
\centering\includegraphics[]{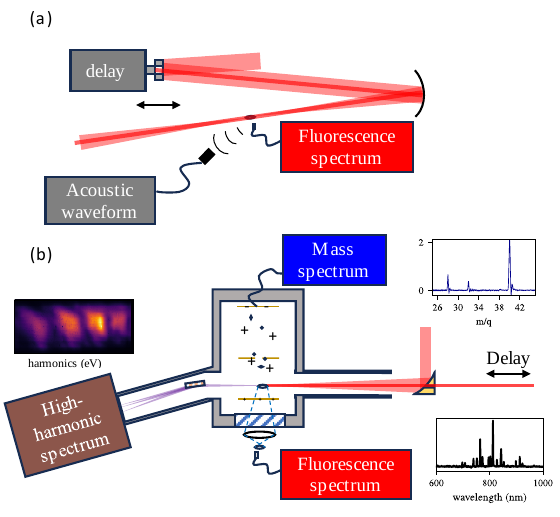}
\caption{\label{fig:schemeoptica} \textbf{(a)} Sketch of the setup for measurements in ambient air, using fluorescence and acoustic signals. \textbf{(b)} Sketch of the setup for in-situ measurements in the vacuum compartments of the beamline, using \ac{TOF} spectrometry, high-harmonic and fluorescence spectroscopy.}
\end{figure}

\subsection{Generation of octave-spanning few cycle pulses}
Few-cycle laser pulses were created by spectrally broadening 25-fs, 1.6-mJ pulses centered at 800~nm that originated from a titanium:sapphire-based laser system operated at 1040~Hz. For this purpose, a 1.1-m-long hollow-core fiber with an inner diameter of 250~µm was uniformly filled with helium at 1.8~bar. The resulting, spectrally broadened pulses were then compressed using 8 reflections (in total) off two pairs of chirped mirrors (PC70, UltraFast Innovations) and transmission through a pair of fused-silica wedges.

\subsection{Measurements in ambient air}\label{subsub:ambient}
Figure \ref{fig:schemeoptica}a illustrates the simple setup that was used for waveform reconstruction in ambient air. A concentrically split mirror was employed to divide the impinging beam (beam diameter 6.2~mm) into a strong and a weaker pulse with identical waveforms. While the outer annulus (inner diameter 3~mm) corresponded to $\epsilon_{\text{hi}}$, reflection off the inner mirror (outer diameter 1~mm) resulted in $\epsilon_{\text{lo}}$, which could be delayed by means of a linear piezo-electric stage (PI, resolution 0.1~nm). Subsequently, the pulses were focused in air to generate a single plasma filament, using a spherical mirror with 40~cm focal length. The fluorescence from the plasma onset region was then focused with a plano-convex lens (BK7, focal length 75~mm) into a spectrometer (Ocean Optics Maya 2000 Pro) through an optical fiber (Ocean Optics QP1000-2-UV-VIS, core diameter 1~mm). Subsequently, the acoustic waveform was recorded via a commercially available low cost microphone (Knowles EK series model 3033, 1--2~kHz) and an oscilloscope (Teledyne Lecroy Wavejet Touch 354, 500MHz, 2GS/s) in a back-to-back manner. Microphone bandwidth limits and ringing were mitigated by integrating the entire power spectrum of each recorded acoustic waveform. With the split-mirror geometry the relative delay between $\epsilon_{\text{hi}}$ and $\epsilon_{\text{lo}}$ is inherently stable. The entire air-plasma setup fits on a compact $18\times18$~cm$^2$ breadboard (with margin), representing a small, portable and inexpensive pulse characterization unit. Paired with lazyCRIME's fast, accurate reconstruction, the setup's portability and versatility enables routine waveform-resolving pulse characterization within minutes and compatibility with a wide range of optical layouts.

\subsection{In-situ measurements in vacuum}\label{subsub:vacuum}
Figure \ref{fig:schemeoptica}b depicts the relevant components of the measurement setup inside the \ac{ATAS} experimental apparatus. This setup was chosen to exemplify the variety of approaches that can be followed to implement in-situ waveform reconstruction at most comparable endstations without intrusion into the usual acquisition procedure. Three modes of detection were used that each focused on a different nonlinear light-matter response: ion yields after strong-field ionization, fluorescence yield from strong-field excitation and high-harmonic yield from \ac{HHG}.

After coupling into the vacuum compartments of the beamline, two collinear beams of $\epsilon_{\text{hi}}$ and $\epsilon_{\text{lo}}$ were created by splitting and subsequent recombination with two holey parabolic mirrors. The delay between $\epsilon_{\text{hi}}$ and $\epsilon_{\text{lo}}$ was controlled via two piezoelectric stages with interferometric stabilization~\cite{Huppert:15}, resulting in a delay uncertainty of less than 20 as (FWHM). For both \ac{HHG} and fluorescence measurements, pure target gases were delivered by a pulsed Even-Lavie valve\cite{Even:15} with 280~µm nozzle diameter, operating at stagnation pressures between 10 and 17 bar at repetition rates of 260--520~Hz, while residual traces of air in the vacuum chamber (\ce{H2O}, \ce{N2}, \ce{O2}, \ce{Ar}) were used as a target for the ion yield measurements. Ion yields were measured by means of \ac{TOF} mass spectrometry using an electrode array in Eppink-Parker design \cite{Eppink:97} and a common microchannel plate detector. The high-harmonic spectra were recorded in the same way as during transient absorption measurements, using a concave flat-field grating (Hitachi 001-0660, 50--248~eV) and a low-noise camera (Princeton Instruments, PIXIS-XO 2KB)). Fluorescence spectra were recorded with a spectrometer (Ocean Optics Maya 2000 Pro) located outside of the vacuum chamber, by collecting the plasma emission through a BK7 window with a plano-convex lens (BK7, 75~mm focal length) and focusing it into an optical fiber (Ocean Optics QP1000-2-UV-VIS, core diameter 1mm). We would like to draw the reader's attention in particular to this fluorescence-based detection scheme, since it represents the by far least perturbing and most inexpensive detection scheme. The only requirement for its implementation is that $\epsilon_{\text{hi}}$ creates a sufficiently intense plasma fluorescence that is visible through a vacuum window.

\section{lazyCRIME -- Fast and simple \ac{CRIME}-based pulse reconstruction algorithm}

\begin{figure}
\centering\includegraphics[]{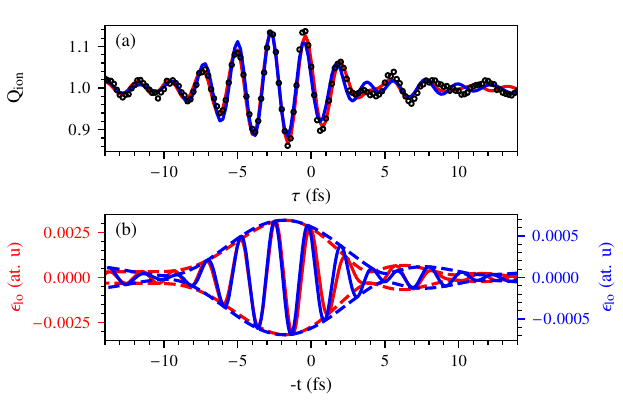}
\caption{\label{fig:N2CRIMEvsLC} \textbf{(a)} Measured (black dots) and simulated relative yield of $\mathrm{N}_2^+$ ion using CRIME (red) and lazyCRIME (blue) reconstructed waveform. \textbf{(b)} Reconstructed field $\epsilon_\text{lo}$ using CRIME and $\epsilon_\text{lo}$ from lazyCRIME, respectively, including their envelopes (dashed lines). The time axis is inverted to illustrate similarity between $\epsilon_\text{lo}$ and $Q_\text{ion}$.}
\end{figure}

\ac{CRIME} provides complete reconstruction of the waveforms of both involved fields on an absolute scale. However, it requires the spectra and peak fluences of both pulses as input in addition to the relative signal trace. We hereby introduce a novel simplified version of \ac{CRIME}, which we dub lazyCRIME. In contrast to \ac{CRIME}, lazyCRIME requires only the relative signal trace and a single spectrum on arbitrary scale. Our new algorithm thus enables straightforward and easy waveform characterization, even when the focus profiles of the two beams are not accessible, or they cannot be spatially separated to measure their individual spectra. lazyCRIME assumes that the waveforms of the two laser fields are equivalent in shape, making it less versatile than \ac{CRIME}, but easier to use in practice. Upon successful parameter optimization, the user will obtain a laser-electric field trace with only coarse absolute field strength scaling but an accurately shaped waveform, allowing for a precise analysis of the pulse's spectral and temporal properties. For example, Fig. \ref{fig:N2CRIMEvsLC}b shows, using the same relative yield measurement, both reconstructions obtained with \ac{CRIME} and LazyCRIME are in excellent agreement. The expected minor pedestal deviation in the \ac{CRIME} and LazyCRIME reconstruction arises from the waveform mismatch between $\epsilon_{\text{hi}}$ and $\epsilon_{\text{lo}}$ that is implied by their difference in spectra (Fig. \ref{fig:N2}a,b).

In contrast to the \ac{TIPTOE} retrieval algorithm~\cite{Cho:21}, the application of lazyCRIME is, in principle, not limited by the intensity ratio between $\epsilon_\text{hi}$ and $\epsilon_\text{lo}$. The inclusion of a measured spectrum as input furthermore ensures that the lazyCRIME-reconstructed field is free of unphysical frequency components, for example, from experimental noise.

In addition to a single set of spectral phases, which, along with a scalar spectral phase offset, describes the two fields equally, the lazyCRIME algorithm also determines two additional parameters, $F_\text{hi}$ and $q_F$. While $F_\text{hi}$ brings the peak fluence of the stronger pulse to absolute scale, $q_F$ sets the peak fluence ratio between $\epsilon_\text{hi}$ and $\epsilon_\text{lo}$. Consequently, the parameter space for a lazyCRIME optimization is just above half as large as that for \ac{CRIME}, resulting in severely accelerated convergence. The reduction in dimensionality allows the algorithm to run on a standard desktop computer in minutes. For example, the reconstruction in Fig. \ref{fig:acoustic}(c) was obtained in less than two minutes on an 18-core Intel(R) Core(TM) i9-10900 CPU.

In this work \ac{ADK} tunneling rate \cite{Ammosov:86,Tong:05} is the model of choice to describe the nonlinear response during CRIME reconstruction. We would like to emphasize that the CRIME algorithm and its variants, by their modular nature, can be easily adapted to other physical models of choice.

\section{Waveform reconstruction using different nonlinear responses}
In this section we present back-to-back measurements based on pairs of different nonlinear observables, each for the same target species and identical laser conditions, and compare the resulting waveforms from \ac{CRIME} or lazyCRIME, as appropriate.

\subsection{High-harmonic versus ion yield} \label{subsect:HHG}

\begin{figure}
\centering\includegraphics[]{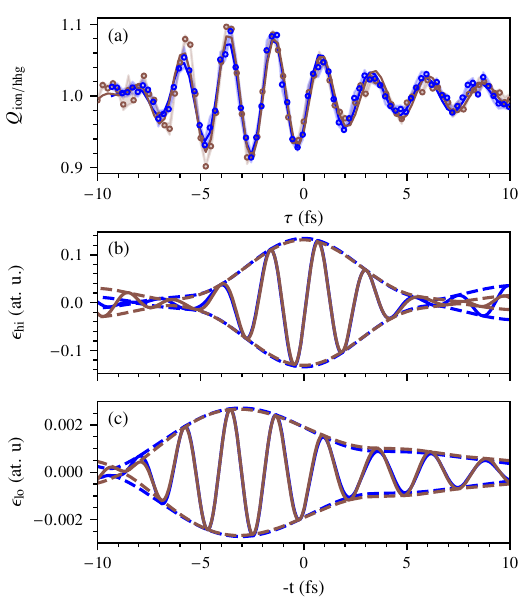}
\caption{\label{fig:Ar}  \textbf{(a)} Measured (dot) and simulated (solid line) relative yields $Q$ of \ce{Ar+} ions (blue) and total high harmonics from \ce{Ar} (brown). Shaded areas: one standard deviation. \textbf{(b), (c)} Reconstructed electric fields with envelopes (dashed).}
\end{figure}

First, we compare waveform measurements using \ce{Ar+} ion yields and the high-harmonic yield from \ce{Ar}, which was assessed by integrating over several harmonic orders easily observable in our setup. We would like to stress here that until now, only the combination of ion yield measurements and \ac{CRIME} has been shown to allow waveform reconstruction with absolute electric field strengths~\cite{Wiese:24}.

Figure \ref{fig:Ar}a shows the measured relative yields $Q_\text{exp}$ for the two cases. They agree very well in both amplitude and shape. Waveform reconstruction was performed with the \ac{CRIME} algorithm, mapping the spectral exposures of $\epsilon_{\text{hi}}$ and $\epsilon_{\text{lo}}$ each onto 20 frequency bands. The results are shown in Fig.~\ref{fig:Ar}b,c. For both $\epsilon_{\text{hi}}$ and $\epsilon_{\text{lo}}$, the resulting waveforms are in excellent agreement, demonstrating the suitability of the total high-harmonic yield for retrieving absolute laser-electric fields with waveform resolution. The \acs{HHG}-based reconstruction scheme represents a method that is by far the easiest to implement at common \ac{ATAS} endstations, because it shares sample delivery and detector with the measurement scheme for transient absorption. At our beamline, switching from transient absorption to \acs{HHG}-based waveform measurement only requires the removal of a metallic filter that usually blocks residual visible laser light along the path of the extreme-ultraviolet beam.

\subsection{Fluorescence versus ion yield}\label{subsect:N2}

\begin{figure}
\centering\includegraphics[]{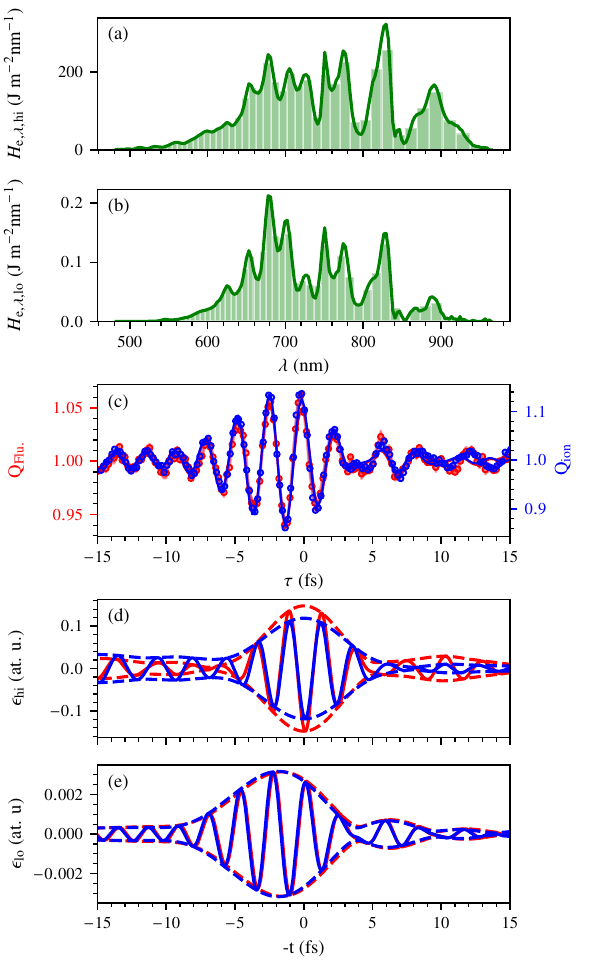}
\caption{\label{fig:N2}\textbf{(a),(b)} Measured $H_\mathrm{e,\lambda}$ (green) of $\epsilon_{\text{hi}}$ and $\epsilon_{\text{lo}}$ with $N_{\text{hi}} = N_{\text{lo}} = 40$ frequency bands (green bars) for \ac{CRIME} reconstruction. \textbf{(c)} Measured (dot) and simulated (solid line) relative signal yield from integrated plasma fluorescence $Q_\mathrm{Flu}$ (red) and $\mathrm{N}_2^+$ ion yield $Q_\mathrm{ion}$ (blue). Shaded areas: one standard deviation. \textbf{(d),(e)} Reconstructed $\epsilon_{\text{hi}}$ and $\epsilon_{\text{lo}}$ with envelopes (dashed).}
\end{figure}

Let us now look at the usability of plasma fluorescence for waveform retrieval, again contrasting it to an equivalent ion yield measurement. Figure \ref{fig:N2}c shows the measured relative yields of total fluorescence of \ce{N2} (spectrally integrated from 300--500~nm) and \ce{N2+} ionization yield. The relative ion and fluorescence yields match perfectly in terms of shape but differ in modulation depth. This indicates that the two underlying nonlinear phenomena---strong-field ionization and excitation of fluorescing states---depend differently on electric field strength. The reduced contrast in the relative fluorescence yield may arise from several processes---for example from photon re-scattering/attenuation in the plasma or population redistribution during post-ionization dynamics. However, identifying the dominant mechanism is not trivial~\cite{Zhang:21,Xu:15} and lies beyond the scope of this study. Nevertheless, the \ac{CRIME} algorithm outputs waveform pairs (see Fig.\ref{fig:N2}d,e) that are in good agreement for the two detection methods. For $\epsilon_{\text{lo}}$, which contains the most precise information of the two reconstructed waveforms, the agreement is virtually perfect. Merely the results for $\epsilon_{\text{hi}}$ show a deviation in terms of absolute field strength, which is most likely rooted in the inadequate use of the \ac{ADK} tunneling rate to model the fluorescence yield. Therefore through \ac{CRIME}, total fluorescence can also be used to retrieve absolute field strength for $\epsilon_{\text{lo}}$ and, with reduced fidelity, for $\epsilon_{\text{hi}}$. 

\subsection{Sound intensity versus Fluorescence yield}\label{subsect:acoustic}

\begin{figure}
\centering\includegraphics[]{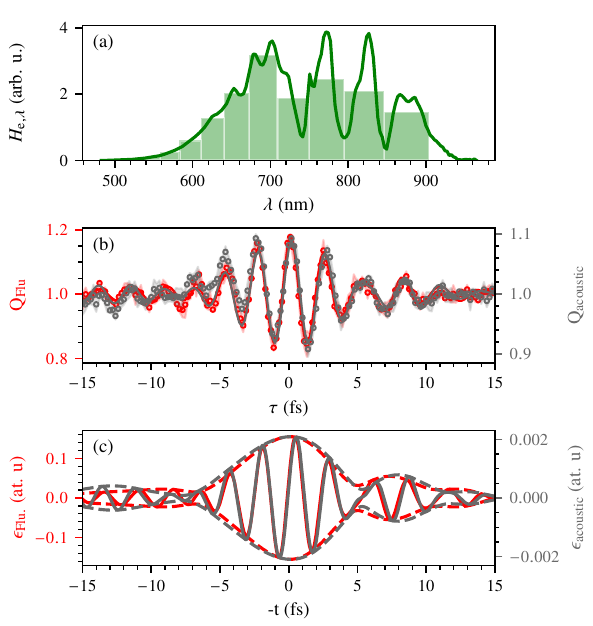}
\caption{\label{fig:acoustic}\textbf{(a)} Measured $H_\mathrm{e,\lambda}$ (green) of the combined driving field and its mapping to 10 frequency bands for lazyCRIME. \textbf{(b)} Measured (dot) and simulated (solid line) relative signal yield $Q$ from integrated plasma fluorescence (red) integrated acoustic power spectrum (grey). Shaded: one standard deviation. \textbf{(c)} Reconstructed fields from fluorescence (red) and acoustic (grey) relative yields with envelopes (dashed).}
\end{figure}

In the following, we will contrast a waveform measurement using the sound intensity emitted from a plasma with one relying on plasma fluorescence. Both measurements were conducted in ambient air, using the setup described in \ref{subsub:ambient}. Figure \ref{fig:acoustic}b shows the measured relative yields from total fluorescence and from the integrated acoustic power spectrum. The two relative yield traces agree well in terms of shape, though the acoustic trace exhibits a slight baseline shift near a delay of $-5$~fs. Such a distortion has already been reported for an acoustic measurement in ambient air, where it manifested as a low-frequency component~\cite{Han:25}. We observed similar baseline distortions when we collected the fluorescence from the end of the plasma filament instead of the beginning, or when the plasma-driving intensity was further increased. This suggests that the source of the distortion is rooted in the plasma retroacting on the light field.

However, such baseline distortions can be easily mitigated by reducing the peak intensity in the focus.

For this experimental setup using the split-mirror geometry, it is not as straightforward to measure the focal profiles and spectra of $\epsilon_{\text{hi}}$ and $\epsilon_{\text{lo}}$ individually. We hence utilized the lazyCRIME algorithm for the waveform retrieval, since it requires only a single spectrum on arbitrary scale as input. As can be seen in Fig.~\ref{fig:acoustic}b,c, the two detection methods result in nearly identical simulated relative yield and electric waveforms. The reconstruction procedure proved insensitive towards both low-frequency distortions and high-frequency noise in the measured trace. So also the sound emitted from a plasma represents a usable observable for accurate waveform retrieval, which is consistent with an earlier study that employed laser light centered at 900~nm at a repetition rate of 2~kHz~\cite{Han:25}.

\section{Waveform reconstruction using different molecular targets}\label{subsect:Fmol}

\begin{figure}
\centering\includegraphics[]{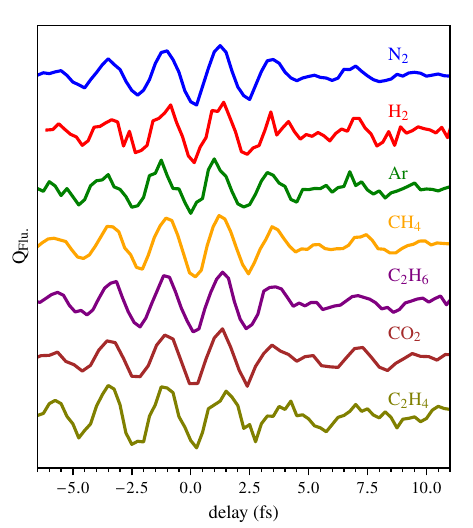}
\caption{\label{fig:Fmol}Measurements of the relative integrated fluorescence yield from $\mathrm{N}_2$ (blue), $\mathrm{H}_2$ (red), Ar (green), $\mathrm{CH}_4$ (orange), $\mathrm{C}_2\mathrm{H}_6$ (purple), $\mathrm{CO}_2$ (brown) and \ce{C2H4}(olive). Traces are arbitrarily rescaled and vertically shifted for visual comparison.}
\end{figure}

\begin{figure}
\centering\includegraphics[]{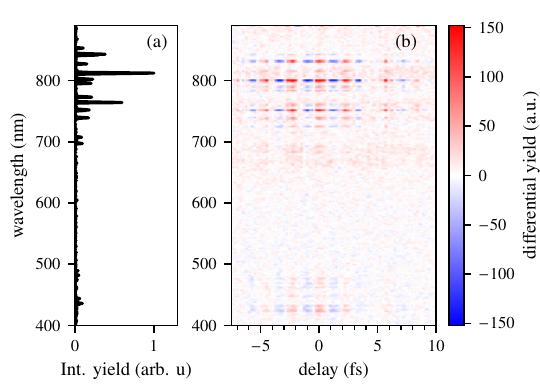}
\caption{\label{fig:Ar_2d}\textbf{(a)} Integrated fluorescence spectrum from $\mathrm{Ar}$. \textbf{(b)} Delay-dependent differential fluorescence spectrum of $\mathrm{Ar}$, obtained by subtracting the reference fluorescence spectrum with $\epsilon_{\text{hi}}$ only}
\end{figure}

Let us now look at how the choice of the target species influences the retrievable waveform information. For this purpose, we compare the relative total fluorescence yields of argon and various gas-phase molecules. All measurements were performed in vacuum, following the setup delineated in \ref{subsub:vacuum}. Fig.~\ref{fig:Fmol} shows the recorded relative yield traces for \ce{Ar}, \ce{N2}, \ce{H2}, \ce{CH4} (methane), \ce{C2H6} (ethane), \ce{CO2} (carbon dioxide) and \ce{C2H4} (ethylene). Within the experimental uncertainty, there is no discernible difference in the shape of the traces. So, at least for the given laser pulse properties, waveform reconstruction can be performed regardless of the sample target for relative fluorescence yield measurement.

For each species, we extracted the relative fluorescence yield by integrating all observed fluorescence lines. Within experimental uncertainty, the temporal structure retrieved was independent of which lines were included. For example, Fig.~\ref{fig:Ar_2d} shows the delay-dependent differential fluorescence spectrum of \ce{Ar}. All observed fluorescence lines exhibit the same modulation as a function of delay. We found the same to be true for all investigated molecular targets. Not only does this finding justify the use of the spectrally integrated fluorescence yield, it also provides rationalization for utilizing a photodiode instead of a spectrometer to collect the total fluorescence, which would make the fluorescence detection scheme even more inexpensive and easier to implement. In conjunction with its above-shown independence of the fluorophore and the fact that it can be implemented without any perturbation of the common workflow of the apparatus, the fluorescence-based detection scheme is an outstanding approach for in-situ waveform retrieval.

\section{Conclusion}
In this work, we have performed measurements of numerous common non-linear observables -- ion yield, fluorescence yield, sound intensity and high harmonic yield -- across a plethora of atom/molecules with distinctive properties. In all cases, highly consistent and reliable waveform reconstructions were obtained by means of the CRIME/lazyCRIME algorithm. This comprehensive experimental study shows that relative yield measurements using any non-linear response or any medium will in general give equivalent waveform reconstructions, with the temporal resolution following just from the degree at which the delay can be controlled. Leveraging this medium independence, we showed minimally intrusive, in-situ characterization schemes that easily integrate into typical workflows of ultrafast spectroscopy. In particular, we demonstrated that in-situ waveform characterization can be performed completely \textit{non-intrusively} at an apparatus that is commonly used for \ac{ATAS} experiments. We would like to highlight, in particular, the use of plasma fluorescence for this purpose because it can be collected without any intrusion into the common workflow, for example, by using a photodiode.

By pairing with \ac{CRIME} and the newly introduced reduced-input variant lazyCRIME, robust waveform reconstruction can be performed on a standard desktop computer within minutes. These capabilities enable fast, accurate, and resilient waveform-resolved characterization across virtually all conventional contexts of ultrafast spectroscopy. In sum, this work advanced the frontier of ultrafast science by introducing a general, economical framework for rapid, routine and highly adaptable waveform characterization that is accessible to a broad community of scientists in ultrafast spectroscopy.

\section{Back matter}

\begin{backmatter}
\bmsection{Funding}
ETH Zürich, Swiss National Science Foundation (grant 200020\_204928)

\bmsection{Acknowledgment}
We thank A. Schneider and M. Seiler for their technical support, T. Balciunas and T. Kopp for establishing the acquisition software, and J. Bredehoeft for general support in the experimental setup. 

\bmsection{Disclosures}
Chatgpt-5 was used only for grammatical proofreading purpose during the drafting process.

\noindent The authors declare no conflicts of interest.

\bmsection{Data Availability Statement}\label{sec:dataavail}
The source codes for CRIME and lazyCRIME are openly available on \url{https://github.com/jimicurlyhead/crime}.

\end{backmatter}


\bibliography{sample}

@article{kopp25a,
  title={Field-resolved measurements of soliton-self-compressed single-cycle pulses and their application to water-window high-harmonic generation},
  author={Kopp, Tristan and Redaelli, Leonardo and Wiese, Joss and Fazio, Giuseppe and others},
  journal={{Optica}, accepted}
}

@article{lanfaloni25a,
  title={Self-compressed waveform-stable light transients enabling water-window attosecond spectroscopy},
  author={Lanflaoni, Valentina Utrio and Vismarra, Federico and Ardali, Emir and Monahan, Nicholas and others},
  journal={Nature Photonics, accepted}
}

@article{gaumnitz17a,
  title={Streaking of 43-attosecond soft-X-ray pulses generated by a passively CEP-stable mid-infrared driver},
  author={Gaumnitz, Thomas and Jain, Arohi and Pertot, Yoann and Huppert, Martin and Jordan, Inga and Ardana-Lamas, Fernando and W{\"o}rner, Hans Jakob},
  journal={Optics express},
  volume={25},
  number={22},
  pages={27506--27518},
  year={2017},
  publisher={Optical Society of America}
}

@article{Matselyukh:25b,
  title={Petahertz Polarimetry Using Tunneling Ionization},
  author={Matselyukh, Danylo T and Cho, Wosik and Kim, Bin and M{\"u}ller, Gian and W{\"o}rner, Hans Jakob and Kim, Kyung Taec},
  journal={Ultrafast Science},
  volume={5},
  pages={0113},
  year={2025},
  publisher={AAAS}
}

@article{Wiese:24,
author = {Joss Wiese and Katherine Brupbacher and Jona Bredehoeft and Danylo T. Matselyukh and Hans Jakob W\"orner},
journal = {Opt. Express},
number = {27},
pages = {48734--48747},
publisher = {Optica Publishing Group},
title = {Universal and waveform-resolving dual pulse reconstruction through interferometric strong-field ionization},
volume = {32},
month = {Dec},
year = {2024},
url = {https://opg.optica.org/oe/fulltext.cfm?uri=oe-32-27-48734&id=566031},
doi = {10.1364/OE.534553},
}

@article{Han:25,
author = {Meng Han and Ming-Chang Chen and Ming-Shian Tsai and Hao Liang},
journal = {Optica},
number = {4},
pages = {459--464},
publisher = {Optica Publishing Group},
title = {Hearing carrier-envelope offset frequency and phase in air with a microphone},
volume = {12},
month = {Apr},
year = {2025},
url = {https://opg.optica.org/optica/fulltext.cfm?uri=optica-12-4-459&id=569715},
doi = {10.1364/OPTICA.545857},
}

@article{Zhang:21,
author = {Chaojie Zhang and Zan Nie and Yipeng Wu and  Mitchell Sinclair and Chen-Kang Huang and Ken A Marsh and Chan Joshi},
journal = {Plasma Phys. Control. Fusion},
number = {9},
pages = {095003},
publisher = {IOP Publishing},
title = {Ionization induced plasma grating and its applications in strong-field ionization measurements},
volume = {63},
month = {Aug},
year = {2021},
url = {https://iopscience.iop.org/article/10.1088/1361-6587/ac1751},
doi = {10.1088/1361-6587/ac1751},
}

@article{Xu:15,
author = {Huailiang Xu and Erik L{\"o}tstedt and Atsushi Iwasaki and Kaoru Yamanouchi},
journal = {Nat. Commun.},
pages = {8347},
publisher = {Nature Publishing Group},
title = {Sub-10-fs population inversion in N$_2^+$ in air lasing through multiple state coupling},
volume = {6},
month = {Sep},
year = {2015},
url = {https://doi.org/10.1038/ncomms9347},
doi = {10.1038/ncomms9347},
}

@article{Vismarra:24,
author = {Federico Vismarra and Marina Fern{\'a}ndez-Gal{\'a}n and Daniele Mocci and Lorenzo Colaizzi and V{\'i}ctor Wilfried Segundo and Roberto Boyero-Garc{\'i}a and Javier Serrano and Enrique Conejero-Jarque and Marta Pini and Lorenzo Mai and Yingxuan Wu and Hans Jakob W{\"o}rner and Elisa Appi and Cord L. Arnold and Maurizio Reduzzi and Matteo Lucchini and Julio San Rom{\'a}n and Mauro Nisoli and Carlos Hern{\'a}ndez-Garc{\'i}a and Roc{\'i}o Borrego-Varillas},
journal = {Light Sci. Appl.},
pages = {197},
publisher = {Nature Publishing Group},
title = {Isolated attosecond pulse generation in a semi-infinite gas cell driven by time-gated phase matching},
volume = {13},
month = {Aug},
year = {2024},
url = {https://doi.org/10.1038/s41377-024-01564-5},
doi = {10.1038/s41377-024-01564-5},
}

@article{Matselyukh:25,
author = {Danylo Matselyukh and V{\'\i}t Svoboda and Hans Jakob W{\"o}rner},
journal = {Nat. Commun.},
pages = {6540},
publisher = {Nature Publishing Group},
title = {Attosecond X-ray spectroscopy reveals the competing stochastic and ballistic dynamics of a bifurcating Jahn--Teller dissociation},
volume = {16},
month = {Jul},
year = {2025},
url = {https://doi.org/10.1038/s41467-025-61512-8},
doi = {10.1038/s41467-025-61512-8},
}

@article{Niederhausen:08,
  author = {Thomas Niederhausen and Uwe Thumm},
  journal = {Phys. Rev. A},
  number = {1},
  pages = {013407},
  publisher = {American Physical Society},
  title = {Controlled vibrational quenching of nuclear wave packets in D$_2^+$},
  volume = {77},
  month = {Jan},
  year = {2008},
  doi = {10.1103/PhysRevA.77.013407},
}

@article{Uiberacker:07,
author = {Markus Uiberacker and Tobias Uphues and Martin Schultze and Alexander J. Verhoef and Vladislav Yakovlev and Matthias F. Kling and Joachim Rauschenberger and Nils M. Kabachnik and Hans Schr{\"o}der and Matthias Lezius and K. L. Kompa and Helmut G. Muller and Marc J. J. Vrakking and Ferenc Krausz and Ursula Kleineberg},
journal = {Nature},
number = {7136},
pages = {627--632},
publisher = {Nature Publishing Group},
title = {Attosecond real-time observation of electron tunnelling in atoms},
volume = {446},
month = {Apr},
year = {2007},
doi = {10.1038/nature05648},
url = {https://doi.org/10.1038/nature05648},
}

@article{Kluender:11,
author = {K. Kl{\"u}nder and J. M. Dahlstr{\"o}m and M. Gisselbrecht and T. Fordell and M. Swoboda and D. Gu{\'e}not and P. Johnsson and J. Caillat and J. Mauritsson and A. Maquet and R. Ta{\"i}eb and A. L'Huillier},
journal = {Phys. Rev. Lett.},
number = {14},
pages = {143002},
publisher = {American Physical Society},
title = {Probing single-photon ionization on the attosecond time scale},
volume = {106},
month = {Apr},
year = {2011},
doi = {10.1103/PhysRevLett.106.143002},
url = {https://doi.org/10.1103/PhysRevLett.106.143002},
}

@article{Hentschel:01,
author = {M. Hentschel and R. Kienberger and C. Spielmann and G. A. Reider and N. Milosevic and T. Brabec and P. Corkum and U. Heinzmann and M. Drescher and F. Krausz},
journal = {Nature},
pages = {509--513},
publisher = {Nature Publishing Group},
title = {Attosecond metrology},
volume = {414},
month = {Nov},
year = {2001},
doi = {10.1038/35107000},
url = {https://doi.org/10.1038/35107000},
}

@article{Kraus:15,
author = {Peter M. Kraus and Beno\^{i}t Mignolet and Denitsa Baykusheva and Andrea Rupenyan and L{\'e}o Horn{\'y} and Evgeny F. Penka and Gopal Dixit and Francesco A. Lima and Rolf Cirelli and Stephanie Zeller and Markus Lucchini and Martin Vacher and Joachim Ullrich and Fabrice Remacle and Hans Jakob W{\"o}rner},
journal = {Science},
number = {6262},
pages = {790--795},
publisher = {American Association for the Advancement of Science},
title = {Measurement and laser control of attosecond charge migration in ionized iodoacetylene},
volume = {350},
month = {Nov},
year = {2015},
doi = {10.1126/science.aab2160},
url = {https://doi.org/10.1126/science.aab2160},
}

@incollection{Baykusheva:20,
author = {Denitsa Baykusheva and Hans Jakob W{\"o}rner},
title = {Attosecond Molecular Spectroscopy and Dynamics},
booktitle = {Molecular Spectroscopy and Quantum Dynamics},
publisher = {Elsevier},
year = {2020},
chapter = {Attosecond Molecular Spectroscopy and Dynamics},
}

@article{Wong:94,
author = {Victor Wong and Ian A. Walmsley},
journal = {Opt. Lett.},
number = {4},
pages = {287--289},
publisher = {Optica Publishing Group},
title = {Analysis of ultrashort pulse-shape measurement using linear interferometers},
volume = {19},
month = {Feb},
year = {1994},
doi = {10.1364/OL.19.000287},
url = {https://doi.org/10.1364/OL.19.000287},
}

@article{Amat-Roldan:05,
author = {Ivan Amat-Rold{\'a}n and Iain G. Cormack and Pablo Loza-Alvarez and David Artigas},
journal = {Opt. Lett.},
number = {9},
pages = {1063--1065},
publisher = {Optica Publishing Group},
title = {Measurement of electric field by interferometric spectral trace observation},
volume = {30},
month = {May},
year = {2005},
doi = {10.1364/OL.30.001063},
url = {https://doi.org/10.1364/OL.30.001063},
}

@article{Kane:93,
author = {D. J. Kane and R. Trebino},
journal = {IEEE J. Quantum Electron.},
number = {2},
pages = {571--579},
publisher = {IEEE},
title = {Characterization of arbitrary femtosecond pulses using frequency-resolved optical gating},
volume = {29},
month = {Feb},
year = {1993},
doi = {10.1109/3.199311},
url = {https://doi.org/10.1109/3.199311},
}

@article{Miranda:12,
author = {Miguel Miranda and Thomas Fordell and Cord Arnold and Anne L'Huillier and Helder Crespo},
journal = {Opt. Express},
number = {1},
pages = {688--697},
publisher = {Optica Publishing Group},
title = {Simultaneous compression and characterization of ultrashort laser pulses using chirped mirrors and glass wedges},
volume = {20},
month = {Jan},
year = {2012},
doi = {10.1364/OE.20.000688},
url = {https://doi.org/10.1364/OE.20.000688},
}

@article{Keusters:03,
author = {Dorine Keusters and Howe-Siang Tan and Patrick O'Shea and Erik Zeek and Rick Trebino and Warren S. Warren},
journal = {J. Opt. Soc. Am. B},
number = {10},
pages = {2226--2237},
publisher = {Optica Publishing Group},
title = {Relative-phase ambiguities in measurements of ultrashort pulses with well-separated multiple frequency components},
volume = {20},
month = {Oct},
year = {2003},
doi = {10.1364/JOSAB.20.002226},
url = {https://doi.org/10.1364/JOSAB.20.002226},
}

@book{Keller:22,
author = {Ursula Keller},
title = {Ultrafast Lasers: A Comprehensive Introduction to Fundamental Principles with Practical Applications},
publisher = {Springer},
address = {Cham},
series = {Graduate Texts in Physics},
year = {2022},
doi = {10.1007/978-3-030-82532-4},
url = {https://link.springer.com/book/10.1007/978-3-030-82532-4},
isbn = {978-3-030-82531-7},
}

@article{Gallmann:00,
author = {L. Gallmann and D. H. Sutter and N. Matuschek and G. Steinmeyer and U. Keller},
journal = {Appl. Phys. B},
number = {Suppl 1},
pages = {S67--S75},
publisher = {Springer},
title = {Techniques for the characterization of sub-10-fs optical pulses: a comparison},
volume = {70},
month = {Jun},
year = {2000},
doi = {10.1007/s003400000307},
url = {https://link.springer.com/article/10.1007/s003400000307},
}

@article{Cho:21,
	title = {Reconstruction algorithm for tunneling ionization with a perturbation for the time-domain observation of an electric-field},
	volume = {11},
	issn = {2045-2322},
	url = {https://www.nature.com/articles/s41598-021-92454-y},
	doi = {10.1038/s41598-021-92454-y},
	language = {en},
	number = {1},
	urldate = {2024-05-28},
	journal = {Scientific Reports},
	author = {Cho, Wosik and Shin, Jeong-uk and Kim, Kyung Taec},
	month = jun,
	year = {2021},
	pages = {13014},
}

@article{Even:15,
author = {U. Even},
journal = {EPJ Techniques and Instrumentation},
pages = {17},
publisher = {Springer},
title = {The Even{-}Lavie valve as a source for high intensity supersonic beam},
volume = {2},
year = {2015},
doi = {10.1140/epjti/s40485-015-0027-5},
url = {https://doi.org/10.1140/epjti/s40485-015-0027-5},
}

@article{Eppink:97,
author = {Andr{\'e} T. J. B. Eppink and David H. Parker},
journal = {Rev. Sci. Instrum.},
number = {9},
pages = {3477--3484},
publisher = {AIP Publishing},
title = {Velocity map imaging of ions and electrons using electrostatic lenses: Application in photoelectron and photofragment ion imaging of molecular oxygen},
volume = {68},
month = {Sep},
year = {1997},
doi = {10.1063/1.1148310},
url = {https://doi.org/10.1063/1.1148310},
}

@article{Huppert:15,
author = {M. Huppert and I. Jordan and H. J. W{\"o}rner},
journal = {Rev. Sci. Instrum.},
number = {12},
pages = {123106},
publisher = {AIP Publishing},
title = {Attosecond beamline with actively stabilized and spatially separated beam paths},
volume = {86},
month = {Dec},
year = {2015},
doi = {10.1063/1.4937623},
url = {https://doi.org/10.1063/1.4937623},
}

@article{Itatani:02,
author = {J. Itatani and F. Qu{\'e}r{\'e} and G. L. Yudin and M. Yu. Ivanov and F. Krausz and P. B. Corkum},
journal = {Phys. Rev. Lett.},
number = {17},
pages = {173903},
publisher = {American Physical Society},
title = {Attosecond Streak Camera},
volume = {88},
month = {Apr},
year = {2002},
doi = {10.1103/PhysRevLett.88.173903},
url = {https://doi.org/10.1103/PhysRevLett.88.173903},
}

@article{Goulielmakis:04,
author = {E. Goulielmakis and M. Uiberacker and R. Kienberger and A. Baltuska and V. Yakovlev and A. Scrinzi and Th. Westerwalbesloh and U. Kleineberg and U. Heinzmann and M. Drescher and F. Krausz},
journal = {Science},
number = {5688},
pages = {1267--1269},
publisher = {American Association for the Advancement of Science},
title = {Direct Measurement of Light Waves},
volume = {305},
month = {Aug},
year = {2004},
doi = {10.1126/science.1100866},
url = {https://doi.org/10.1126/science.1100866},
}

@article{Park:18,
author = {Seung Beom Park and Kyungseung Kim and Wosik Cho and Sung In Hwang and Igor Ivanov and Chang Hee Nam and Kyung Taec Kim},
journal = {Optica},
number = {4},
pages = {402--408},
publisher = {Optica Publishing Group},
title = {Direct sampling of a light wave in air},
volume = {5},
month = {Apr},
year = {2018},
doi = {10.1364/OPTICA.5.000402},
url = {https://doi.org/10.1364/OPTICA.5.000402},
}

@article{Yeom:24,
author = {Kyunghoon Yeom and Wosik Cho and Jeong-Uk Shin and Bin Kim and Sung In Hwang and Jae Hee Sung and Kyung Taec Kim},
journal = {Opt. Express},
number = {13},
pages = {23796--23802},
publisher = {Optica Publishing Group},
title = {Single-shot measurement of a laser waveform using plasma fluorescence in ambient air},
volume = {32},
month = {Jun},
year = {2024},
doi = {10.1364/OE.527805},
url = {https://opg.optica.org/oe/abstract.cfm?URI=oe-32-13-23796},
}

@article{Heinzerling:25,
author = {Amelie M. Heinzerling and Francesco Tani and Manoram Agarwal and Vladislav S. Yakovlev and Ferenc Krausz and Nicholas Karpowicz},
journal = {Nat. Photon.},
pages = {772--777},
publisher = {Nature Publishing Group},
title = {Field-resolved attosecond solitons},
volume = {19},
month = {Jul},
year = {2025},
doi = {10.1038/s41566-025-01658-5},
url = {https://doi.org/10.1038/s41566-025-01658-5},
}

@article{Saito:18,
author = {Nariyuki Saito and Nobuhisa Ishii and Teruto Kanai and Jiro Itatani},
journal = {Opt. Express},
number = {19},
pages = {24591--24601},
publisher = {Optica Publishing Group},
title = {All-optical characterization of the two-dimensional waveform and the Gouy phase of an infrared pulse based on plasma fluorescence of gas},
volume = {26},
year = {2018},
doi = {10.1364/OE.26.024591},
url = {https://doi.org/10.1364/OE.26.024591},
}

@article{Awad:24,
author = {Mohanad Awad and Apurba Manna and Sebastian Hell and Bo Ying and Levente {\'A}br{\'o}k and Zsolt Div{\'e}ki and Eric Cormier and B{\'a}lint Kiss and Jan B{\"o}hmer and Carsten Ronning and Seung Heon Han and Antony George and Andrey Turchanin and Adrian N. Pfeiffer and Matthias K{\"u}bel},
journal = {Opt. Express},
number = {2},
pages = {1325--1333},
publisher = {Optica Publishing Group},
title = {Few-cycle laser pulse characterization on-target using high-harmonic generation from nano-scale solids},
volume = {32},
year = {2024},
doi = {10.1364/OE.508062},
url = {https://doi.org/10.1364/OE.508062},
}

@article{Schotz:22,
author = {J. Sch{\"o}tz and A. Maliakkal and J. Bl{\"o}chl and others},
journal = {Nat. Commun.},
pages = {962},
publisher = {Nature Publishing Group},
title = {The emergence of macroscopic currents in photoconductive sampling of optical fields},
volume = {13},
year = {2022},
doi = {10.1038/s41467-022-28412-7},
url = {https://doi.org/10.1038/s41467-022-28412-7},
}

@article{Schiffrin:13,
author = {A. Schiffrin and T. Paasch-Colberg and N. Karpowicz and V. Apalkov and D. Gerster and S. M{\"u}hlbrandt and M. Korbman and J. Reichert and M. Schultze and S. Holzner and J. V. Barth and R. Kienberger and R. Ernstorfer and V. S. Yakovlev and M. I. Stockman and F. Krausz},
journal = {Nature},
pages = {70--74},
publisher = {Nature Publishing Group},
title = {Optical-field-induced current in dielectrics},
volume = {493},
month = {Jan},
year = {2013},
doi = {10.1038/nature11567},
url = {https://doi.org/10.1038/nature11567},
}

@article{Fan:16,
author = {G. Fan and T. Bal{\v{c}}i{\=u}nas and C. Fourcade-Dutin and S. Haessler and A. A. Voronin and A. M. Zheltikov and F. G{\'e}r{\^o}me and F. Benabid and A. Baltu{\v{s}}ka and T. Witting},
journal = {Opt. Express},
number = {12},
pages = {12713--12729},
publisher = {Optica Publishing Group},
title = {X-SEA-F-SPIDER characterization of over octave spanning pulses in the infrared range},
volume = {24},
month = {Jun},
year = {2016},
doi = {10.1364/OE.24.012713},
url = {https://opg.optica.org/oe/abstract.cfm?URI=oe-24-12-12713},
}

@article{Ammosov:86,
author = {M. V. Ammosov and N. B. Delone and V. P. Krainov},
journal = {Sov. Phys. JETP},
pages = {1191--1194},
publisher = {AIP Publishing},
title = {Tunnel ionization of complex atoms and of atomic ions in an alternating electromagnetic field},
volume = {64},
year = {1986},
}

@article{Tong:05,
author = {X. M. Tong and C. D. Lin},
journal = {J. Phys. B: At. Mol. Opt. Phys.},
number = {15},
pages = {2593--2600},
publisher = {IOP Publishing},
title = {Empirical formula for static field ionization rates of atoms and molecules by lasers in the barrier-suppression regime},
volume = {38},
month = {Jul},
year = {2005},
doi = {10.1088/0953-4075/38/15/001},
url = {https://doi.org/10.1088/0953-4075/38/15/001},
}

\end{document}